\documentclass[prl,twocolumn,superscriptaddress,nobibnotes,letterpaper]{revtex4} 
\usepackage[utf8]{inputenc}
\usepackage[T1]{fontenc}
\usepackage{braket}
\usepackage{amsmath}
\usepackage{caption}
\usepackage{subcaption}

\usepackage{multirow}
\usepackage{graphicx}
\usepackage{xcolor} 

\begin{document}

\title{Deployed MDI-QKD and Bell-State Measurements Coexisting with Standard Internet Data and Networking Equipment}

\author{Remon C. Berrevoets}\thanks{These authors contributed equally to this work.}
\affiliation{QuTech, Delft University of Technology, Delft, The Netherlands}
\author{Thomas Middelburg}\thanks{These authors contributed equally to this work.}
\affiliation{QuTech, Delft University of Technology, Delft, The Netherlands}
\author{Raymond F. L. Vermeulen}
\affiliation{QuTech, Delft University of Technology, Delft, The Netherlands}
\author{Luca Della Chiesa}
\affiliation{Cisco Systems, Inc., 170 West Tasman Dr. San Jose CA, USA}
\author{Federico Broggi}
\affiliation{Cisco Systems, Inc., 170 West Tasman Dr. San Jose CA, USA}
\author{Stefano Piciaccia}
\affiliation{Cisco Systems, Inc., 170 West Tasman Dr. San Jose CA, USA}
\author{Rene Pluis}
\affiliation{Cisco Systems, Inc., 170 West Tasman Dr. San Jose CA, USA}
\author{Prathwiraj Umesh}
\affiliation{QuTech and Kavli Institute of Nanoscience, Delft University of Technology, Delft, The Netherlands}
\author{Jorge F. Marques}
\affiliation{QuTech and Kavli Institute of Nanoscience, Delft University of Technology, Delft, The Netherlands}
\author{Wolfgang Tittel}
\affiliation{QuTech and Kavli Institute of Nanoscience, Delft University of Technology, Delft, The Netherlands}
\author{Joshua A. Slater}
\email{j.a.slater@tudelft.nl}
\affiliation{QuTech, Delft University of Technology, Delft, The Netherlands}

\begin{abstract}
The forthcoming quantum Internet is poised to allow new applications not possible with the conventional Internet. The ability for both quantum and conventional networking equipment to coexist on the same fiber network would greatly facilitate the deployment and adoption of coming quantum technology. Most quantum networking tasks, like quantum repeaters and the connection of quantum processors, require nodes for multi-qubit quantum measurements (often Bell-State measurements), and their real-world coexistence with the conventional Internet has yet to be shown. Here we field deploy an MDI-QKD system, containing a Bell-State measurement Node, over the same fiber network as multiple standard IP data networks, between three nearby cities in the Netherlands.  We demonstrate over 10~Gb/s data communication rates simultaneously with our next-generation QKD system, and estimate 200~GB/s of classical data transmission would be easily achievable without significantly affecting QKD performance.  Moreover, as the network ran autonomously for two weeks, this shows an important step towards the coexistence and integration of quantum networking into the existing telecommunication infrastructure.
\end{abstract}

\maketitle

\section*{Introduction}
The conventional Internet, the applications that run on it, and the telecommunication networks on which it operates, have had tremendous impact on society.  The coming quantum Internet is poised to have a similar impact by enabling new applications that are fundamentally not possible with the conventional Internet~\cite{wehner_quantum_2018}. Amongst those applications are secure access to remote quantum computers and long-distance networked quantum computers~\cite{pompili2021realization}, enhanced sensors~\cite{ge2018distributed}, enhanced clock synchronization accuracy~\cite{komar2016quantum}, teleportation of quantum information across a network~\cite{bennett1993teleporting}, as well as security-related applications such as information-theoretic secure processing via blind quantum computing~\cite{broadbent2009universal} and perhaps the most-known application, quantum key distribution (QKD) for information-theoretic secure distribution of cryptographic keys~\cite{bennett_quantum_1984,gisin_quantum_2002,lo_secure_2014}.  With ever increasing attention from industry and governments on these developments, many academic efforts have turned towards integration of quantum communication technologies with conventional networking technology. Fundamentally, this means quantum optic signals – e.g.~single photons level pulses – and conventional networking signals and data – e.g.~telecom bright laser light – sharing the same optical fiber. In other words, at the physical layer~\cite{zimmermann1980osi}, there is a desire for quantum signals and telecom signals to coexist on the fiber.

Coexistence of quantum signals and telecom signals is challenging, considering the large intensity difference and the general intolerance to noise by quantum receivers and detectors~\cite{townsend1997simultaneous,peters2009dense,chapuran2009optical,eraerds2010quantum,patel2012coexistence,patel_quantum_2014,frohlich2015quantum,aleksic2015perspectives,kumar2015coexistence,wang2017long,frohlich2017long,eriksson2018coexistence,grunenfelder2021limits,alleaume_technology_2020,vokic_deployment_2020}. The natural approach is to separate quantum signals from telecom signals with  well-established wavelength division multiplexing (WDM) techniques. In WDM fiber technology, each signal is transmitted at a dedicated wavelength channel and off-the-shelf components can well multiplex the different wavelengths onto a single fiber, and then demultiplex the channels at the receiver station, directing them to their matched receivers. The difficulty of combining quantum signals and telecom signals on the same fiber, even when employing WDM, comes from spurious noise photons and crosstalk that the bright telecom light will inevitably add to the quantum channel. The primary mechanism by which this occurs is Raman scattering~\cite{chapuran2009optical}; an inelastic process wherein photon-phonon interactions scatter light within a wide range of wavelengths around the bright telecom light, including, potentially, into the quantum dedicated wavelength channel. In addition to WDM devices, ameliorating the effects of Raman scattering can be achieved by the addition of narrow spectral filters, temporal filters, and by maintaining a large wavelength separation between the quantum and telecom channels.

Numerous quantum communication experiments have employed these techniques to demonstrate coexistence of quantum signals with conventional network data on the same fiber~\cite{townsend1997simultaneous,peters2009dense,chapuran2009optical,eraerds2010quantum,patel2012coexistence,patel_quantum_2014,frohlich2015quantum,aleksic2015perspectives,kumar2015coexistence,wang2017long,frohlich2017long,eriksson2018coexistence,grunenfelder2021limits,alleaume_technology_2020,vokic_deployment_2020}, however all have been demonstrations of prepare-and-measure (P\&M) QKD, in which a transmitting network node (Alice) prepares a qubit state and transmits it to a receiving node (Bob) for detection. A drawback to these studies is that P\&M quantum communication, while suitable for trusted node QKD~\cite{gisin_quantum_2002,lo_secure_2014}, does not include important ingredients for future stages of the quantum Internet~\cite{wehner_quantum_2018}, e.g. multi-photon interference measurements such as Bell-state measurements (BSMs). Such BSM stations – often referred to as midpoints, heralding stations, or Center Nodes – play a critical role in quantum teleportation~\cite{bennett1993teleporting}, entangling quantum processors, linking distantly separated quantum computers~\cite{pompili2021realization}, quantum repeaters~\cite{briegel1998quantum}, and next-generation QKD systems and networks~\cite{lo_measurement-device-independent_2012,braunstein_side-channel-free_2012,pile2018twin,liu2019experimentalTF}.

Measurement-Device-Independent (MDI) QKD is based on such BSMs and can therefore be seen as a stepping stone towards the quantum internet~\cite{lo_measurement-device-independent_2012,braunstein_side-channel-free_2012}. In MDI-QKD, two parties (Alice and Bob) individually prepare and send qubits to a Center Node at which a Bell State Measurement (BSM) is performed. After the Center Node sends the BSM results back to the two parties, Alice and Bob can form entanglement-like correlations and thus, they can generate a secret key via the usual QKD post-processing techniques. 

MDI-QKD also has other advantages when compared to P\&M QKD systems: By placing all single-photon detectors in a Bell-state measurement at the Center Node, MDI-QKD is inherently protected against all known and yet-to-be-proposed detector side-channel attacks. This is an important advantage because historically, detection side-channels attacks have proven to be easily implementable and difficult to defend against~\cite{lim_random_2015,lu_two-way_2013,alhussein_differential_2019,yin_reference-free-independent_2014,duvsek2000unambiguous,makarov2006effects,zhao2008quantum,makarov2009controlling,lydersen2010thermal,lydersen2010hacking,jain2011device,gerhardt2011full,bugge2014laser}.  Furthermore, MDI-QKD allows for many users to connect to each other via a single Center Node.  This brings a new level of practicality and scalability as it allows for multipoint-to-multipoint functionality and sharing of potentially expensive resources, such as single-photon detectors, which can all be placed within the Center Node.

Over the last decade, numerous realizations of MDI-QKD have been shown, both in the lab and in the field~\cite{rubenok2015mdi,da2013proof,valivarthi2015measurement,yin2016measurement,comandar2016quantum,tang2016measurement, liu2019experimental,woodward_gigahertz_2021}. However, so far only one laboratory study has examined how conventional optical communication signals may impact this protocol~\cite{valivarthi2019measurement}, which guides our development presented here: a field-deployed, multi-node quantum communication system, incorporating a BSM node and coexisting with conventional telecom equipment, signals and data traffic.

In this letter we report the demonstration of MDI-QKD deployed between three cities, wherein its quantum signals coexist in the same optical fiber as two functional telecom IP data networks. At the Center Node, we use various strategies to isolate the quantum signals from the telecom signals, while fully maintaining the data networks, and providing over 10 Gb/s Internet connectivity.  
\section*{Experimental Setup}

\subsection*{The MDI-QKD Protocol}

In the MDI-QKD protocol the two End Nodes (Alice and Bob) are functionally identical: they randomly choose a string of qubit states, each being one of the four BB84 states ($\ket{0}$, $\ket{1}$ in the $Z$-basis and $\ket{\pm} = (\ket{0} \pm \ket{1})/\sqrt{2}$ in the $X$-basis). Alice and Bob associate the states $\ket{0}$ and $\ket{+}$ ($\ket{1}$ and $\ket{-}$) with classical bit value 0 (and 1). They sequentially encode these qubit states into attenuated laser pulses at the single-photon level and transmit them to a Center Node. For each pair of photons (one from Alice and one from Bob, which arrive simultaneously at the Center Node) the Center Node performs a Bell-state measurement (BSM), which may project the qubits onto the maximally entangled $\ket{\psi-}$ Bell state.  After each attempted BSM, the Center Node immediately announces to the End Nodes whether the BSM was successful or not, and Alice and Bob only store information about the states of the created qubits that resulted in a BSM. After a sufficiently large number of BSMs, the End Nodes move to the standard QKD post-processing phase.  Note that our implementation described below does not use perfect single photons, but instead weak coherent laser pulses. To protect against the threat of the photon number splitting attack~\cite{brassard2000limitations}, our End Nodes also randomly choose between three mean photon numbers for each pulse (referred to as signal, decoy, and vacuum) and employ a three-intensity decoy state analysis to analyze their data~\cite{yu_three-intensity_2013}, which allows secure distribution of key, even without a true single photon source.

For post-processing, Alice and Bob use an authenticated classical channel to first perform basis reconciliation and discard data about qubit pairs for which they have selected different bases. Of the remaining qubit pairs, they reveal a subset of the bit values so as to estimate for each basis the error rate as well as the probability of a projection onto the $\ket{\psi-}$ Bell state per emitted qubit pair (known as the gain).  They use the data from the $X$-basis to bound information an eavesdropper could have learned about the key during photon transmission. Then, to finally distill a secret key from their data, they perform classical error correction on the $Z$-basis data and privacy amplification to remove the number of bits of information that could have been leaked to an eavesdropper. This results in a secret key rate:
\begin{equation}
    \label{eqn:secretkey}
	R = [s_{11}^Z[1-H(e_{11}^X)]-Q_{ss}^{Z}fH(E_{ss}^Z)],
\end{equation}
where R is the secure key rate per pair of Z-basis signal intensity qubits sent, $s_{11}^Z$ is the single photon gain of the Z-basis,  $e_{11}^X$ is the single photon  error rate in the X-basis, both extracted from the decoy analysis~\cite{yu_three-intensity_2013}, $Q_{ss}^{Z}$ and $E_{ss}^Z$ are the gain and error rate of the Z-basis signal qubits, H is the binary entropy function, and f is the error correction efficiency (set to 1.12 in this work).

\subsection*{The MDI-QKD System}

\begin{figure*}[!ht]
    \centering
    \includegraphics{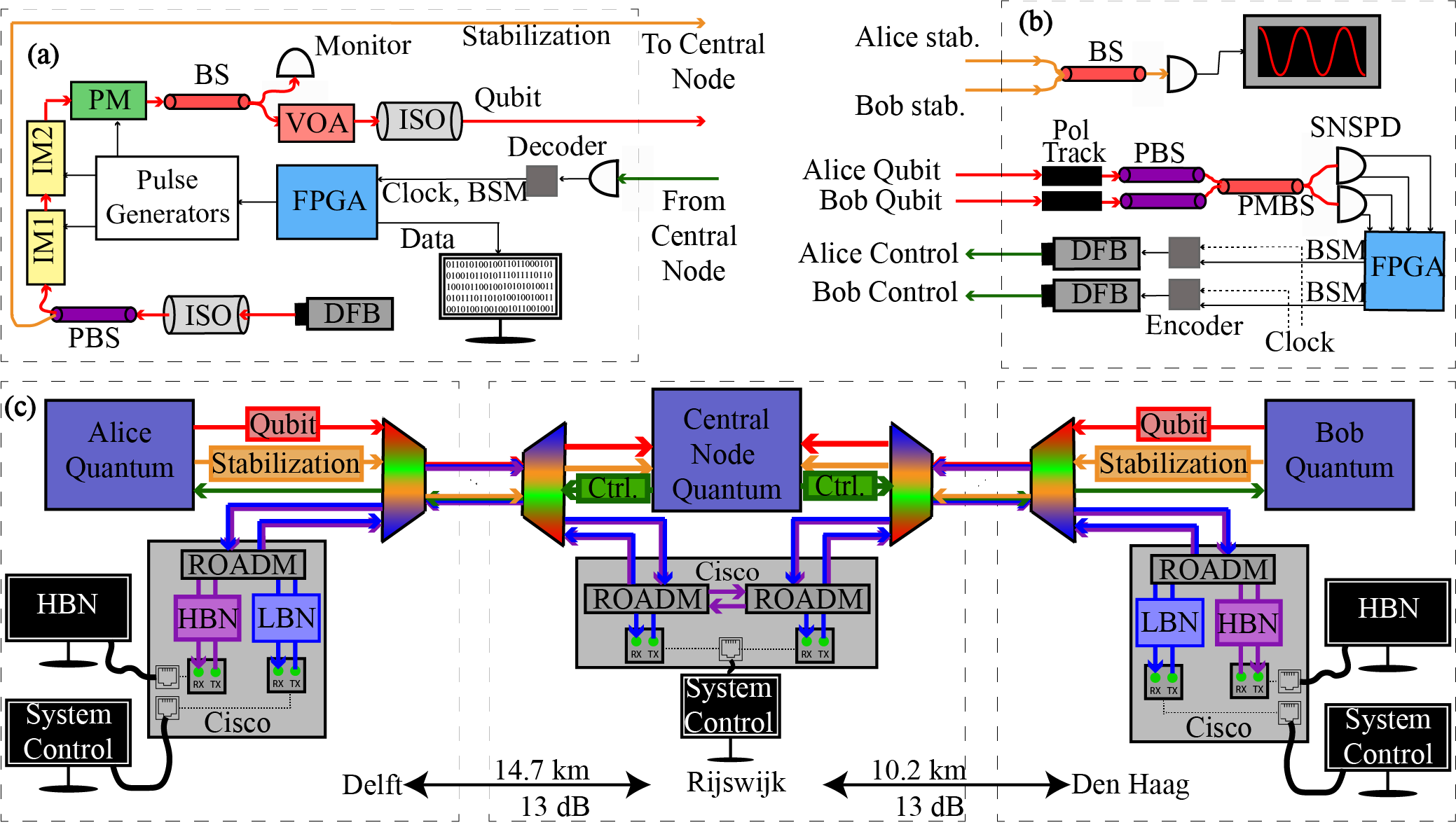}
    \caption{\textbf{(a)} Alice and Bob use 1310~nm wavelength DFB lasers to generate the light for their time-bin qubit states. After the laser, an isolator prevents any light moving back into the laser and a polarizing beam splitter (PBS) ensures the light is well polarized. One of the PBS' outputs is used to generate the time-bin qubits via a series of intensity and phase modulators. The first intensity modulator (IM1) creates optical pulses with a full width at half maximum (FWHM) of 600~ps, separated by 1.5~ns. A phase modulator modulates one of the pulses with a $\pi$-phase to create the $\ket{-}$ state and a second intensity modulator (IM2) creates the different mean photon numbers required for the decoy states. The pulses then pass a variable optical attenuator (VOA) to attenuate all pulses to the single photon level. \textbf{(b)} Qubits interfere at the Center Node on a polarization maintaining beam splitter (PMBS), each output of which is connected to a superconducting nanowire single photon detector (SNSPD) for detection. Digitizer boards with 400-ps wide temporal-filtering windows convert the analog response of the SNSPD into digital signals corresponding to single projections onto qubit states $\ket{early}$ and $\ket{late}$. The Center Node uses two DFB lasers at 1548~nm wavelength to communicate to Alice and Bob this detection data. \textbf{(c)} Schematic drawing of the optical network and WDM scheme used to demonstrate MDI-QKD coexisting with the two conventional IP data channels (HBN and LBN). The qubits are co-propagating with the optical fields of the IP data channels. Details described in main text.}
    \label{fig:experimental_setup}
\end{figure*}

A schematic of our End Nodes is shown in Fig.~\ref{fig:experimental_setup}a, and described in the caption. Our Alice and Bob use 1310~nm wavelength DFB lasers to generate the light for the time-bin qubit states. They create the states $\ket{early}$, $\ket{late}$ (associated with qubit states $\ket{0}$, $\ket{1}$) and $\ket{\pm}$, as well as various mean photon numbers required in the decoy state protocol using a series of intensity and phase modulators.

At the Center Node, incoming qubit pulses interfere on a beam splitter (PMBS) and are detected by single photon detectors. Successful BSMs are identified by an FPGA, which monitors when the two detectors after the PMBS clicked during the same qubit pulse, but with opposite values (i.e.~one detector registering $\ket{early}$ and the other $\ket{late}$). This correspond to a projection onto the entangled $\ket{\psi-}$ Bell state, which identifies a successful Bell state measurement for the MDI-QKD protocol.  During QKD operation, the Center Node uses two DFB lasers at 1548~nm wavelength to immediately inform Alice and Bob of a successful BSM detection. We refer to this communication channel as the Control channel. At Alice and Bob, detection data are processed on FPGAs, with data of interest written to disk. The full schematics of the Center Node's detection setup is shown in Fig.~\ref{fig:experimental_setup}b.

An important requirement for multi-photon interference such as BSMs, is that input photons must be indistinguishable in all degrees of freedom at the PMBS. Our system accomplishes this through a variety of control and stabilization systems. To synchronize the system between the three locations, a master clock signal of 200 MHz is generated at the Center Node and optically sent to Alice and Bob via the Control channel. Thus, the Control channel transmits both the clock signal and detection data described above, which are modulated together into the optical field for communication to the End Nodes.  Alice and Bob demodulate out the clock signal and distribute it to their FPGAs and pulse generators. Alice's and Bob's optical pulses are aligned to the Center Node's digitizer windows with a precision of 10 ps.

To ensure Alice’s and Bob's qubits are indistinguishable in polarization, their qubits pass through PBSs at the Center Node. Preceding the PBSs are electronic polarization trackers that maximize the transmission through the PBS.  To maintain frequency indistinguishability, Alice and Bob use temperature stabilized DFB lasers. A PBS directly in front of the laser (see Fig.~\ref{fig:experimental_setup}) taps a portion of this light, which the End Nodes then send to the Center Node, and at which it interferes on a beamsplitter~\cite{rubenok2015mdi,valivarthi2015measurement}. This light effectively acts as another communication channel, which we refer to as the Stabilization channel. Its interference is registered as intensity beating on a photodiode; the frequency of which allows the Center Node to generate a feedback signal to send to Alice and Bob. Alice and Bob use temperature controllers to minimize their frequency difference to below 25~MHz, which is sufficiently close for two-photon quantum interference~\cite{hong1987measurement} and a minimal relative phase mismatch between X-basis qubits with our set temporal mode spacing.

To estimate the performance of the system in various scenarios, we characterized the quantum state of the emitted qubits, as well as the Center Node's detection system. The qubits are characterized by two parameters~$(m,\phi)$:
\begin{equation} \label{eq:appqubitmodel}
    \ket{\psi} = \sqrt{m}\ket{0} + e^{i\phi}\sqrt{1-m}\ket{1},
\end{equation}
with $0 \leq m \leq 1$. Each qubit state from each End Node has distinct parameters, which are listed in Table~\ref{tab:qubit-params}. The Center Node's detection system is characterized by its detection efficiency, dark counts (i.e.~noise), and the two-photon quantum interference visibility of the Bell-state measurement. While the $m$'s, and dark counts were directly measured, the visibility as well as the $\ket{-}$ state phases were acquired through fitting measured gains and QBERs.

\subsection*{Coexistence with conventional data channels}

To demonstrate the coexistence of MDI-QKD with conventional internet technology, we integrated the MDI-QKD system with a variety of network equipment from Cisco such as the ASR9000 and the NCS5500 routers, along with an optical platform such as the NCS2000. Specifically, all three nodes employed an NCS2006 system with 20-FS-SMR ROADM linecards to multiplex conventional WDM optical traffic. Alice and Bob had an ASR9000 and an NCS5500 aggregation service router respectively, as well as an additional WSE linecard in their NCS2006 systems.

The Cisco routers were configured to provide two IP networks with 10~Gb/s and 100~Mb/s respectively, which we refer to as the high-bandwidth network (HBN) and lower-bandwidth network (LBN). In all demonstrations discussed below, the MDI-QKD system nodes used the LBN to communicate with each other, while the HBN was used for other data unrelated to the operation of the quantum systems or the experiments; e.g., other users’ data, video calls, video streaming etc.

The HBN and LBN optical signals were multiplexed on the same optical fiber as the qubits from the MDI-QKD system, but using different WDM channels.  We constructed the following networks: the LBN was generated by the NCS2006 systems and operated at an out-of-band OSC wavelength of 1510~nm. The IP-level HBN 10~Gb/s service was generated by the ASR9000 and NCS5500 and the optical carrier signals were generated by the WSE linecards in the NCS2006s, and operated in the C-band at 1550.12~nm wavelength. At the Center Node, the HBN was optically amplified and optically routed from one End Node to the other, while the LBN was converted to copper ethernet and re-generated. Both HBN and LBN were set to equal launch powers throughout all experiments. The full network optical multiplexing design is shown in Fig.~\ref{fig:experimental_setup}c.

We designed the networks such that they operated over two fibers between each End Node and the Center Node. These fibers were named fiber~1 and fiber~2 for both End-Node-to-Center-Node links. Fiber~1 was used for Alice's and Bob's transmissions (Tx) of the IP networks optical signals and qubits (Center Node reception, Rx). We chose this co-propagation configuration to minimize scattered light at the single photon detectors of the Center Node~\cite{valivarthi2019measurement}. The qubit channel generated by the MDI-QKD system operated at 1310~nm wavelength and multiplexing the qubits with the IP data signals was achieved by a standard WDM multiplexer. Demultiplexing the qubits at the Center Node was more challenging.  We used a high isolation WDM (>50~dB) to remove as much IP data signal light as possible, after which we used a narrow-band (2~nm) filter with 45~dB isolation on the qubit line to filter out remaining Raman scattered light around the qubit's wavelength. As will be shown below, these two elements provided sufficient spectral filtering to protect the qubit channel from unwanted noise.

On fiber~2 we set the Center Node Tx (Alice/Bob Rx), for the LBN, HBN and MDI Control channel (the latter at 1548~nm wavelength). Furthermore, the Stabilization channel light at 1310~nm wavelength was placed on fiber~2, due to its spectral indistinguishability with respect to the qubits. These signals were all multiplexed and demultiplexed with standard WDMs, as all of the channels on fiber~2 were comparatively tolerant to noise.

\section*{Results}

For our first tests of MDI-QKD coexisting with multiple classical IP data channels, we ran the entire network in our laboratory. In the lab, each End Node was separated from the Center Node by 20~km of spooled fiber with intrinsic loss of 10.5~dB and 9.0~dB at 1310~nm wavelength respectively. In these experiments, the launch power of the IP networks was adjusted such that the received power per channel was about 500~nW; close to the minimal required received power at the Center Node such that both IP networks were running with 100\%~uptime.

\begin{figure}[!t]
    \begin{center}
    \includegraphics[trim={0.3cm 0.3cm 0.3cm 0.3cm},clip,width=0.45\textwidth]{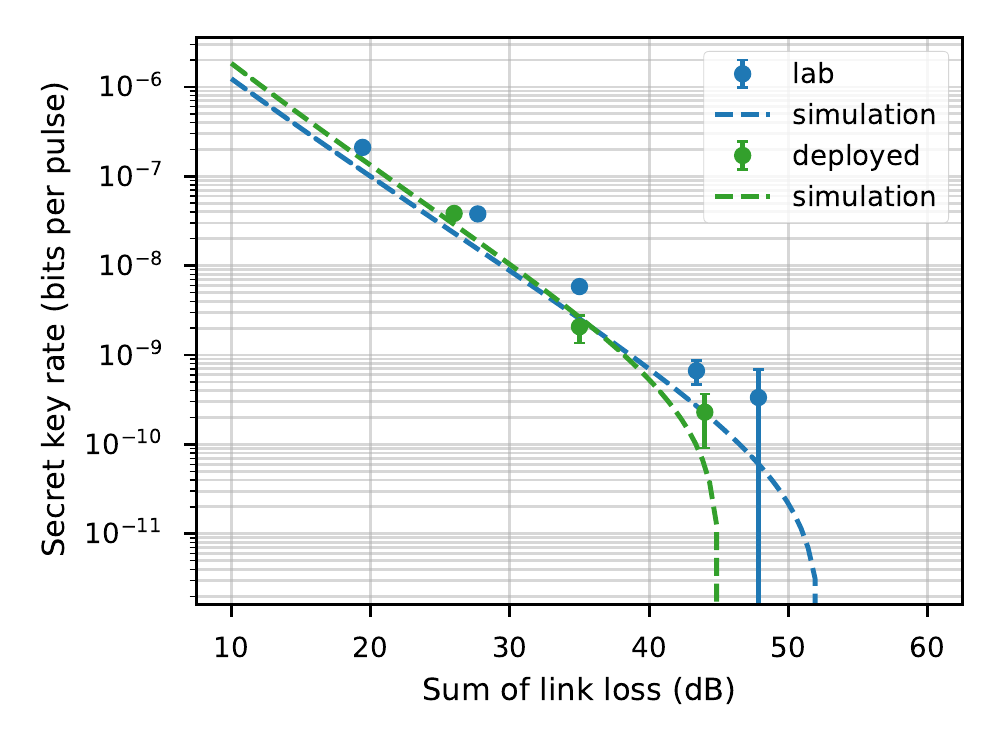}
    \caption{
    Secret key rate as a function of the total link loss between End Nodes. All data was collected while the HBN (10 Gb/s) and LBN (100 Mb/s) IP data channels were present on the same fiber as the qubits. The IP data channels' received power was kept constant at 500 nW for all measurements.  Points are measured data with uncertainty bars representing 1-standard deviation. Dashed lines are simulated results generated from a full characterization of the QKD system.}
    \label{fig:dataset2}
    \end{center}
\end{figure}

We first tested the performance of the network for various attenuations in the fiber network by adding fixed attenuators. In general, performance of decoy-state QKD varies not just with fiber attenuation, but also with the mean photon number of the signal and decoy states. We thus selected, for both End Nodes, mean photon numbers that optimized secret key rate at high attenuation and used these values for all lab experiments.  For all experiments we calculated the expected key rate in the asymptotic regime using equation~\ref{eqn:secretkey}~\cite{yu_three-intensity_2013}.  These results are displayed as the blue points and curve in Fig.~\ref{fig:dataset2}. The losses and resulting gains, QBERs and secret key rates can be found in Table~\ref{tab:gains-qbers-X} and~\ref{tab:gains-qbers-Z}. We found that even with the two coexisting data channels, the system performance would be sufficient for QKD key generation over a large parameter regime: Up to about 50~dB fiber attenuation, corresponding to about 250~km of spooled fiber.

\begin{figure}[!t]
    \begin{center}
    \includegraphics[trim={0.3cm 0.6cm 0.3cm 0.5cm},clip,width=0.5\textwidth]{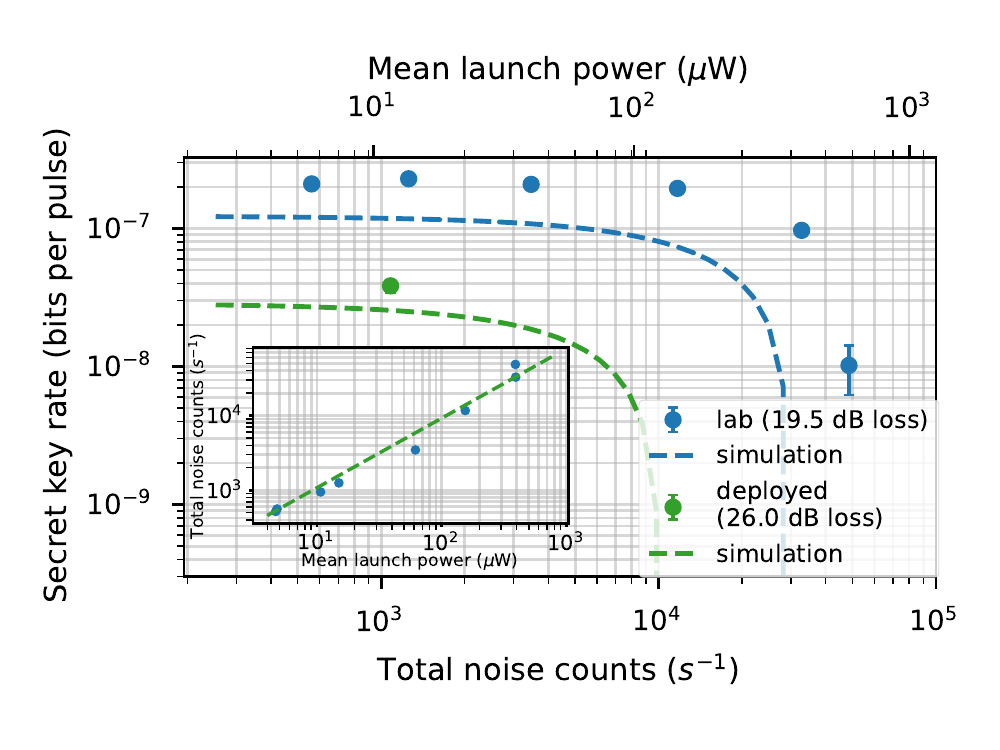}
    \caption{Secret key rates for increasing IP data network launch power (top x-axis) and corresponding noise counts per second on the BSM detectors (bottom x-axis). The smallest launch powers correspond to the minimally required received power at the Center Node for the 2 IP data networks. Inset shows background noise counts per second, summed over all detector windows, as a function of IP data networks' launch power (average launch power from a single End Node).}
    \label{fig:dataset1}
    \end{center} 
\end{figure}

\begin{figure*}[!!!!!!!!!!!!!!!!!!ht]
    \centering
    \includegraphics[width=0.7\textwidth]{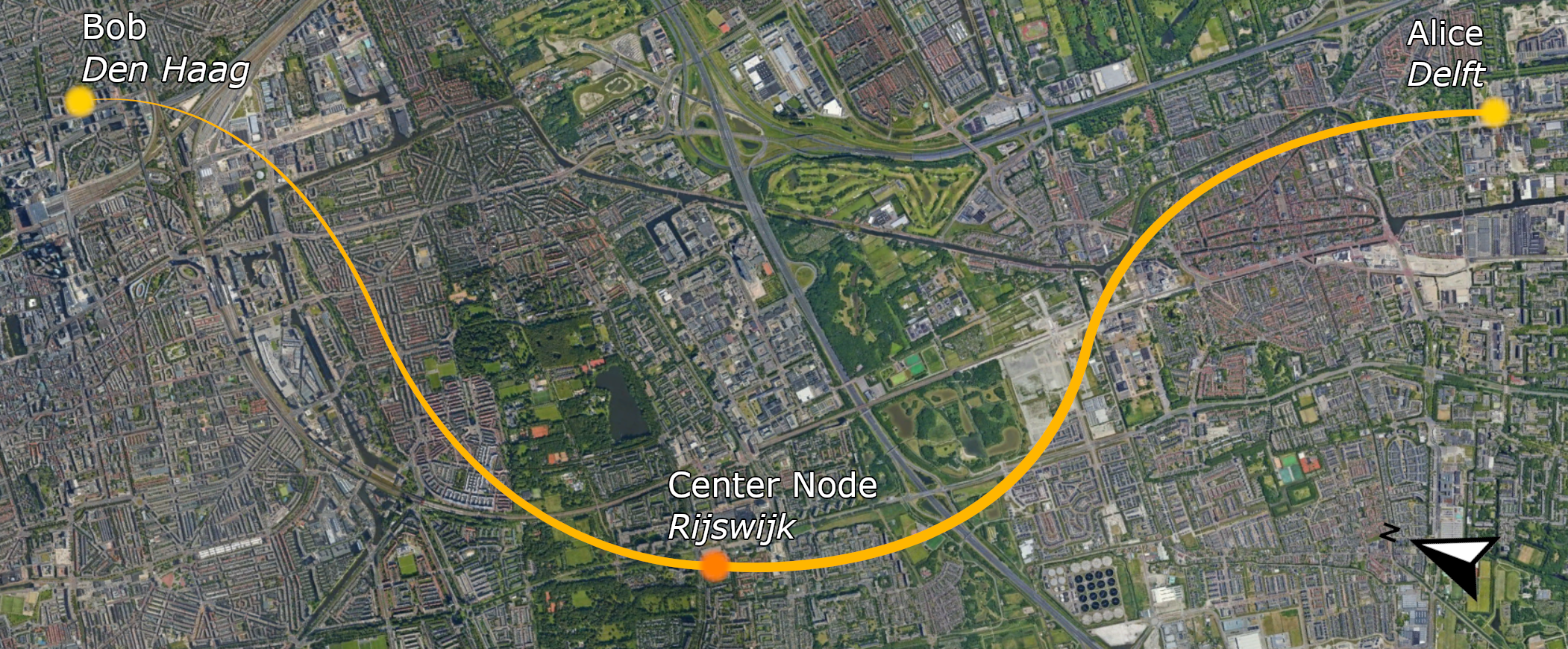}
    \caption{Satellite image showing the locations of the deployed MDI-QKD system. Map data: Google Earth.}
    \label{fig:Deployed_map}
\end{figure*}

In our second set of experiments, we sought to explore the number of simultaneous conventional IP data channels that can coexist on the same fiber as the QKD system without causing significant performance degradation. Effectively, we tested the performance of the QKD network for various launch powers of the IP data channels. In general, as shown in the inset of Fig.~\ref{fig:dataset1}, we found a linear relationship between data launch power and noise on the Center Node's single photon detectors (i.e.~clicks in the absence of qubits), indicating that there exists some remaining crosstalk and/or Raman scattering from the data channels into the qubit channels. Thus, adding further data channels should, at some level, degrade the performance of the QKD channel.  The asymptotic secret key rate for various launch powers can be seen in the blue points and curve of Fig.~\ref{fig:dataset1}. The far-left point (3.5~$\mu$W launch power per channel) on the plot is the initial configuration in which the two data channels (HBC and LBC) have the minimally required received power at the Center Node. The losses and resulting gains, QBERs and secret key rates can be found in Table~\ref{tab:gains-qbers-X-launch} and~\ref{tab:gains-qbers-Z-launch}. We generally observe that increasing the launch power initially has little effect on the MDI-QKD system performance.  In fact, 150~$\mu$W launch power (corresponding to 42 WDM channels) had nearly zero impact, and even pushing to 400~$\mu$W (corresponding to 114 WDM channels) decreased the secret key rate by only a factor of two. This shows a strong resilience of the QKD system to the coexisting data channels. Furthermore, assuming a metropolitan-distance network with the same loss as the fiber spools (about 20~dB), it shows the possibility of an MDI-QKD system coexisting with potentially a hundred conventional 10~Gb/s data channels.

\subsection*{Deployment}

Finally, we deployed the entire network between three cities in the Netherlands: Alice was situated in Delft (QuTech), Bob was situated in Den Haag (KPN test and release center), and the Center Node was situated in Rijswijk (KPN telco distribution/exchange building). Dedicated fibers were made available from each location to the Center Node in Rijswijk with a length of 14.7~km and 10.2~km from Delft and Den Haag respectively. The fibers' losses were equalized to 13~dB using variable optical attenuators (VOAs), totalling 26~dB loss from Alice to Bob. The system was deployed over a six week period (including setup, measurements, demonstration events, and tear down), and fully operating for two weeks in June 2021. During that time the system ran autonomously, except when the network was changed for new measurements. The longest period of uninterrupted operation was 61 hours.

In the field, we performed the same experiments as during the lab test. The performance of the deployed network for various losses is displayed in Fig.~\ref{fig:dataset2} as the green points and curve. Generally we see that the deployed network performs similar to the lab network. The main difference from the lab setting was that we used optimized mean photon numbers for the deployed fiber losses, meaning that the deployed network performed slightly better at lower losses than the lab network, and the lab network was optimized to tolerate higher loss. Nevertheless, there is good agreement between both data sets, and with the simulated performance. This demonstrates that coexisting MDI-QKD and conventional IP data networks can function well, in the field, at metropolitan distances.

Lastly, in Fig.~\ref{fig:dataset1}, we also show the deployed network’s secret key rate as a function of launch power of the IP data network. While the launch powers of the two IP networks were slightly higher to accommodate for the extra loss, the received power was the same as in the lab. The deployed network has lower secret key rates ($4\text{e-}8$ instead of $2\text{e-}7$ secret key bits per pulse), only because it operated over more loss (26~dB loss instead of 19.5~dB).  Importantly, we see that the secret key rate of the deployed network decreased by the expected amount. This again demonstrates the resilience of the MDI-QKD system and the Center Node's BSM detection unit to coexisting conventional data channels. Moreover, our simulations show that a launch power of 100~$\mu$W, corresponding to twenty 10~Gb/s data channels, would only decrease the secret key rate by a factor of 2.  Said differently, given sufficient telecom transceivers, it would be possible to achieve MDI-QKD coexisting with 200~Gb/s conventional IP data transmission in our 26~dB-loss deployed fiber network.

\section*{Discussion}

While we believe that our networks of 19.5~dB and 26~dB fiber attenuation could tolerate 100 and 26 10-Gb/s channels, respectively, increasing the coexisting data capacities further would require extra improvements.  For instance, data channel filtering could be further optimized with custom built, narrower filters.  Alternatively, the MDI-QKD system key rates could be increased and thus make it more resilient to added noise. 

By using high-end routers provided by Cisco for the routing of IP data channels we have demonstrated the integration of MDI-QKD with conventional networks and networking equipment: an important milestone for next-generation QKD systems.  Furthermore, as our network uses a Center Node facilitating Bell-state measurements with qubits coming from other locations in the network, we thus have evidence that the technology for the coming quantum Internet may be made compatible with conventional networks more easily than thought. This marks an important step to practically useful, and scalable quantum communication networks.

\section*{Acknowledgements}
The authors thank Koninklijke KPN N.V. for hosting the network and providing fiber and technical support, Walter van de Garde and Cisco Cyber Security Netherlands for further support, and David Elkouss for valuable discussions. The authors acknowledge funding through the Netherlands Organization for Scientific Research (NWO) and the European project OpenQKD.

\section*{Author Contributions}
RCB, TM, JAS performed the experiments, and with RFLV built the quantum setup, and with important contributions from all authors, wrote the manuscript. The conventional IP networks design was led by LDC, FB, SP, RP, with inputs from all authors. PU, JFM, WT contributed to the early stages of the work. RCB, TM, LDC, FB, SP, RP, JAS integrated the quantum and IP networks.

\bibliography{References}

\clearpage
\onecolumngrid

\section{Supplemental A: Gains and QBERs} \label{sec:AppendixA}
\begin{table}[ht]
\centering
\caption{Measured X-basis Gains and QBERs for the losses in the experiment, for both lab and deployed environments.}
\label{tab:gains-qbers-X}
\begin{tabular}{ccccccccccccc}
    \multicolumn{12}{c}{\textbf{X-basis Gains and QBERs}} \\ 
    \hline \hline
    { } & \textbf{Channel Loss} & {\textbf{Keyrate}} & {} & {} & \multicolumn{3}{c}{\textbf{Gain, $Q_{\mu_i,\mu_j}^X$}} & {} & {}& \multicolumn{3}{c}{\textbf{QBER, $e_{\mu_i,\mu_j}^X$}} \\
    \cline{3-13} 
    
    {}&{} & {} &{} & {} & {$\mu_{b} \rightarrow$} & {} & {} & {} &{}& {} & {}  \\
    
    {}&{} & {} &{$\mu_{a} \downarrow$}  & {} &   $\textit{\textbf{s}}$ & $\textit{\textbf{d}}$ & $\textit{\textbf{v}}$ & {} &{}& $\textit{\textbf{s}}$ & $\textit{\textbf{d}}$ & $\textit{\textbf{v}}$ \\
        
    {}&{} & {} &{}  & $\textit{\textbf{s}}$  & $4.94\cdot10^{-4}$ & $1.12\cdot10^{-4}$  & $9.51 \cdot 10^{-5}$ & {} & $\textit{\textbf{s}}$ & 29.5\%  & 43.9\% & 50\%  \\
    
    {}&{\textbf{19 dB}} & {$2.10\cdot10^{-7}$} &{}  & $\textit{\textbf{d}}$  & $1.75\cdot10^{-4}$ & $2.32\cdot10^{-6}$  & $4.81\cdot10^{-7}$ & {} & $\textit{\textbf{d}}$ & 46.5\% & 32.1\% & 50\% \\ 
    
    {}&{} & {}& {}  & $\textit{\textbf{v}}$  & $1.60\cdot10^{-4}$ & $8.24\cdot10^{-7}$  & $1.08\cdot10^{-9}$ & {} & $\textit{\textbf{v}}$  & 50\%   & 50\%  & 50\%   \\[0.2cm] 
        
    {}&{} & {}& {}  & {} &   $\textit{\textbf{s}}$ & $\textit{\textbf{d}}$ & $\textit{\textbf{v}}$ & {} &{}& $\textit{\textbf{s}}$ & $\textit{\textbf{d}}$ & $\textit{\textbf{v}}$ \\

    {}&{} & {}& {}  & $\textit{\textbf{s}}$  & $5.54\cdot10^{-5}$ & $1.35\cdot10^{-5}$  & $1.15\cdot10^{-5}$ & {} & $\textit{\textbf{s}}$ & 29.2\%  & 44.1\% & 50\%  \\
    
    {}&{\textbf{28 dB}} & {$3.79\cdot10^{-8}$}& {}  & $\textit{\textbf{d}}$  & $1.85\cdot10^{-5}$ & $2.79\cdot10^{-7}$  & $6.40\cdot10^{-8}$ & {} & $\textit{\textbf{d}}$ & 46.0\% & 31.7\% & 50\% \\ 
    
    {}&{} & {} &{}  & $\textit{\textbf{v}}$  & $1.67\cdot10^{-5}$ & $8.81\cdot10^{-8}$  & $9.07\cdot10^{-11}$ & {} & $\textit{\textbf{v}}$  & 50\%  & 50\%  & 50\%   \\[0.2cm]

    {}&{} & {}& {}  & {} &   $\textit{\textbf{s}}$ & $\textit{\textbf{d}}$ & $\textit{\textbf{v}}$ & {} &{}& $\textit{\textbf{s}}$ & $\textit{\textbf{d}}$ & $\textit{\textbf{v}}$ \\
    
    {}&{} & {} &{}  & $\textit{\textbf{s}}$  & $1.02\cdot10^{-5}$ & $3.09\cdot10^{-5}$  & $2.73\cdot10^{-6}$ & {} & $\textit{\textbf{s}}$ & 29.4\%  & 45.3\% & 50\%  \\ 
    
    {\textbf{Lab}}&{\textbf{35 dB}} & {$5.83\cdot10^{-9}$}& {}  & $\textit{\textbf{d}}$  &  $2.75\cdot10^{-6}$ & $5.05\cdot10^{-8}$  & $1.52\cdot10^-8$ & {} & $\textit{\textbf{d}}$ & 45.1\%  & 31.9\% & 50\% \\ 
    
    {}&{} & {}& {}  & $\textit{\textbf{v}}$  & $2.41\cdot10^{-6}$ & $1.27\cdot10^{-8}$  & $1.67\cdot10^{-11}$ & {} & $\textit{\textbf{v}}$  & 50\%  & 50\%  & 50\% \\[0.2cm]  
        
    {}&{} & {}& {}  & {} &  $\textit{\textbf{s}}$ & $\textit{\textbf{d}}$ & $\textit{\textbf{v}}$ & {} &{}& $\textit{\textbf{s}}$ & $\textit{\textbf{d}}$ & $\textit{\textbf{v}}$ \\
    
    {}&{} & {}& {}  & $\textit{\textbf{s}}$  & $1.54\cdot10^{-6}$ & $3.54\cdot10^{-7}$  & $2.99\cdot10^{-7}$ & {} & $\textit{\textbf{s}}$ & 29.5\%  & 44.1\% & 50\%  \\  
    
    {}&{\textbf{43 dB}} & {$6.66\cdot10^{-10}$}& {}  & $\textit{\textbf{d}}$  & $5.37\cdot10^{-7}$ & $7.8\cdot10^{-9}$  & $1.71\cdot10^{-9}$ & {} & $\textit{\textbf{d}}$ & 46.2\%  & 32.3\%& 50\% \\ 
    
    {}&{} & {}& {}  & $\textit{\textbf{v}}$  & $4.86\cdot10^{-7}$ & $2.54\cdot10^{-9}$  & $8.71\cdot10^{-13}$ & {} & $\textit{\textbf{v}}$  & 50\%   & 50\%  & 50\%   \\[0.2cm]
        
    {}&{} & {}& {}  & {} &   $\textit{\textbf{s}}$ & $\textit{\textbf{d}}$ & $\textit{\textbf{v}}$ & {} &{}& $\textit{\textbf{s}}$ & $\textit{\textbf{d}}$ & $\textit{\textbf{v}}$ \\
    
    {}&{} & {}& {}  & $\textit{\textbf{s}}$  & $5.11\cdot10^{-7}$ & $1.28\cdot10^{-7}$  & $1.11\cdot10^{-7}$ & {} & $\textit{\textbf{s}}$ & 29.4\%  & 44.5\% & 50\%  \\
    
    {}&{\textbf{48 dB}} & {$3.35\cdot10^{-10}$}& {}  & $\textit{\textbf{d}}$  & $1.68\cdot10^{-7}$ & $2.72\cdot10^{-9}$  & $6.57\cdot10^{-10}$ & {} & $\textit{\textbf{d}}$ &  45.9\%   & 32.1\%  & 50\% \\  
    
    {}&{} & {}& {}  & $\textit{\textbf{v}}$  & $1.50\cdot10^{-7}$ & $9.64\cdot10^{-10}$  & $2.31\cdot10^{-12}$ & {} & $\textit{\textbf{v}}$  & 50\%  & 50\%  & 50\%   \\[0.2cm] 
    \hline
    
    {}&{} & {}& {}  & {} &   $\textit{\textbf{s}}$ & $\textit{\textbf{d}}$ & $\textit{\textbf{v}}$ & {} &{}& $\textit{\textbf{s}}$ & $\textit{\textbf{d}}$ & $\textit{\textbf{v}}$ \\
    
    {}&{} & {}& {}  & $\textit{\textbf{s}}$  & $1.72\cdot10^{-4}$ & $5.53\cdot10^{-5}$  & $4.99\cdot10^{-5}$ & {} & $\textit{\textbf{s}}$ & 29.9\%  & 46.2\% & 50\%  \\ 
    
    {}&{\textbf{26 dB}} & {$3.84\cdot10^{-8}$}& {}  & $\textit{\textbf{d}}$  & $4.40\cdot10^{-5}$ & $8.01\cdot10^{-7}$  & $2.53\cdot10^{-7}$ & {} & $\textit{\textbf{d}}$ & 44.6\%  & 32.0\% & 50\% \\ 
    
    {}&{} & {}& {}  & $\textit{\textbf{v}}$  & $3.78\cdot10^{-5}$ & $1.76\cdot10^{-7}$  & $1.28\cdot10^{-10}$ & {} & $\textit{\textbf{v}}$  & 50\%   & 50\%  & 50\% \\[0.2cm]  
        
    {}&{} & {}& {}  & {} &  $\textit{\textbf{s}}$ & $\textit{\textbf{d}}$ & $\textit{\textbf{v}}$ & {} &{}& $\textit{\textbf{s}}$ & $\textit{\textbf{d}}$ & $\textit{\textbf{v}}$ \\
    
    {}&{} & {}& {}  & $\textit{\textbf{s}}$  & $2.25\cdot10^{-5}$ & $7.09\cdot10^{-6}$  & $6.36\cdot10^{-6}$ & {} & $\textit{\textbf{s}}$ & 29.8\%  & 46.0\% & 50\%  \\  
    
    {\textbf{Deployed}}&{\textbf{35 dB}} & {$2.07\cdot10^{-9}$}& {}  & $\textit{\textbf{d}}$  & $5.80\cdot10^{-6}$ & $1.02\cdot10^{-7}$  & $2.99\cdot10^{-8}$ & {} & $\textit{\textbf{d}}$ & 44.8\%  & 32.3\% & 50\% \\ 

    {}&{} & {}& {}  & $\textit{\textbf{v}}$  & $5.02\cdot10^{-6}$ & $2.41\cdot10^{-8}$  & $8.59\cdot10^{-12}$ & {} & $\textit{\textbf{v}}$  & 50\%   & 50\%  &50\%   \\[0.2cm]
        
    {}&{} & {}& {}  & {} &   $\textit{\textbf{s}}$ & $\textit{\textbf{d}}$ & $\textit{\textbf{v}}$ & {} &{}& $\textit{\textbf{s}}$ & $\textit{\textbf{d}}$ & $\textit{\textbf{v}}$ \\
    
    {}&{} & {}& {}  & $\textit{\textbf{s}}$  & $3.18\cdot10^{-6}$ & $1.01\cdot10^{-6}$  & $9.06\cdot10^{-7}$ & {} & $\textit{\textbf{s}}$ & 30.6\%  & 46.2\% & 50\%  \\
    
    {}&{\textbf{44 dB}} & {$2.29\cdot10^{-10}$}& {}  & $\textit{\textbf{d}}$  & $8.10\cdot10^{-7}$ & $1.51\cdot10^{-8}$  & $4.85\cdot10^{-9}$ & {} & $\textit{\textbf{d}}$ & 44.7\%  & 32.5\% & 50\% \\ 
    
    {}&{} & {}& {}  & $\textit{\textbf{v}}$  & $6.87\cdot10^{-7}$ & $3.34\cdot10^{-9}$  & $3.99\cdot10^{-12}$ & {} & $\textit{\textbf{v}}$  & 50\%  & 50\%  & 50\%   \\
\end{tabular}
\end{table}

\begin{table*}
\centering
\caption{Measured Z-basis Gains and QBERs for the losses in the experiment, for both lab and deployed environments.}
\label{tab:gains-qbers-Z}
\begin{tabular}{ccccccccccccc}
    \multicolumn{12}{c}{\textbf{Z-basis Gains and QBERs}} \\ \hline \hline
    { } &\multicolumn{1}{c}{\textbf{Channel Loss}} &{\textbf{Keyrate}} &{} & {} & \multicolumn{3}{c}{\textbf{Gain, $Q_{\mu_i,\mu_j}^Z$}} & {} & {}& \multicolumn{3}{c}{\textbf{QBER, $e_{\mu_i,\mu_j}^Z$}} \\ 
    \cline{3-13} 
    {}&{} & {} &{} & {} & {$\mu_{b} \rightarrow$} & {} & {} & {} &{}& {} & {}  \\
    
    {}&{} & {} &{$\mu_{a} \downarrow$}  & {} &   $\textit{\textbf{s}}$ & $\textit{\textbf{d}}$ & $\textit{\textbf{v}}$ & {} &{}& $\textit{\textbf{s}}$ & $\textit{\textbf{d}}$ & $\textit{\textbf{v}}$ \\
        
    {}& {} &{} &{} & $\textit{\textbf{s}}$  & $2.33\cdot10^{-4}$ & $1.68\cdot10^{-5}$  & $1.40 \cdot 10^{-6}$ & {} & $\textit{\textbf{s}}$ & 0.927\%  & 4.54\% & 50\%  \\
    
    {}&{\textbf{19 dB}} & {$2.10\cdot10^{-7}$} &{}  & $\textit{\textbf{d}}$  & $1.60\cdot10^{-5}$ & $1.07\cdot10^{-6}$  & $4.65\cdot10^{-8}$ & {} & $\textit{\textbf{d}}$ & 5.37\% & 5.20\% & 50\% \\ 
    
    {}& {} &{} &{} & $\textit{\textbf{v}}$  & $1.60\cdot10^{-6}$ & $8.24\cdot10^{-7}$  & $6.18\cdot10^{-8}$ & {} & $\textit{\textbf{v}}$  & 50\%   & 50\%  & 50\%   \\[0.2cm]
        
    {}& {} &{} &{} & {} &   $\textit{\textbf{s}}$ & $\textit{\textbf{d}}$ & $\textit{\textbf{v}}$ & {} &{}& $\textit{\textbf{s}}$ & $\textit{\textbf{d}}$ & $\textit{\textbf{v}}$ \\

    {}& {} &{} &{} & $\textit{\textbf{s}}$  & $2.84\cdot10^{-5}$ & $2.02\cdot10^{-6}$  & $1.23\cdot10^{-7}$ & {} & $\textit{\textbf{s}}$ & 0.735\%  & 3.33\% & 50\%  \\
    
    {}&{\textbf{28 dB}} & {$3.79\cdot10^{-8}$} &{}  & $\textit{\textbf{d}}$  & $2.01\cdot10^{-6}$ & $1.33\cdot10^{-7}$  & $4.26\cdot10^{-9}$ & {} & $\textit{\textbf{d}}$ & 4.46\% & 3.65\% & 50\% \\ 
    
    {}& {} &{} &{} & $\textit{\textbf{v}}$  & $1.68\cdot10^{-7}$ & $5.86\cdot10^{-9}$  & $6.28\cdot10^{-11}$ & {} & $\textit{\textbf{v}}$  & 50\%  & 50\%  & 50\%   \\[0.2cm]

    {}& {} &{} &{} & {} &   $\textit{\textbf{s}}$ & $\textit{\textbf{d}}$ & $\textit{\textbf{v}}$ & {} &{}& $\textit{\textbf{s}}$ & $\textit{\textbf{d}}$ & $\textit{\textbf{v}}$ \\
    
    {}& {} &{} &{} & $\textit{\textbf{s}}$  & $5.06\cdot10^{-6}$ & $3.65\cdot10^{-7}$  & $3.18\cdot10^{-8}$ & {} & $\textit{\textbf{s}}$ & 0.878\%  & 4.66\% & 50\%  \\ 
    
    {\textbf{Lab}}& {\textbf{35 dB}} &{$5.83\cdot10^{-9}$} &{} & $\textit{\textbf{d}}$  &  $3.52\cdot10^{-7}$ & $2.40\cdot10^{-8}$  & $9.77\cdot10^{-10}$ & {} & $\textit{\textbf{d}}$ & 4.29\%  & 4.26\% & 50\% \\ 
    
    {}& {} &{} &{} & $\textit{\textbf{v}}$  & $2.55\cdot10^{-8}$ & $1.11\cdot10^{-9}$  & $2.79\cdot10^{-11}$ & {} & $\textit{\textbf{v}}$  & 50\%  & 50\%  & 50\% \\[0.2cm]
        
    {}& {} &{} &{} & {} &  $\textit{\textbf{s}}$ & $\textit{\textbf{d}}$ & $\textit{\textbf{v}}$ & {} &{}& $\textit{\textbf{s}}$ & $\textit{\textbf{d}}$ & $\textit{\textbf{v}}$ \\
    
    {}& {} &{} &{} & $\textit{\textbf{s}}$  & $7.33\cdot10^{-7}$ & $5.29\cdot10^{-8}$  & $4.50\cdot10^{-9}$ & {} & $\textit{\textbf{s}}$ & 0.965\%  & 4.75\% & 50\%  \\  
    
    {}&{\textbf{43 dB}} & {$6.66\cdot10^{-10}$} &{}  & $\textit{\textbf{d}}$  & $5.31\cdot10^{-8}$ & $3.50\cdot10^{-9}$  & $1.85\cdot10^{-10}$ & {} & $\textit{\textbf{d}}$ & 5.53\%  & 5.80\%& 50\% \\ 
    
    {}& {}&{} &{}  & $\textit{\textbf{v}}$  & $5.18\cdot10^{-9}$ & $2.32\cdot10^{-10}$  & $6.16\cdot10^{-12}$ & {} & $\textit{\textbf{v}}$  & 50\%   & 50\%  & 50\%   \\[0.2cm]
        
    {}& {}&{} &{}  & {} &   $\textit{\textbf{s}}$ & $\textit{\textbf{d}}$ & $\textit{\textbf{v}}$ & {} &{}& $\textit{\textbf{s}}$ & $\textit{\textbf{d}}$ & $\textit{\textbf{v}}$ \\
    
    {}& {} &{} &{} & $\textit{\textbf{s}}$  & $2.63\cdot10^{-7}$ & $2.11\cdot10^{-8}$  & $3.39\cdot10^{-9}$ & {} & $\textit{\textbf{s}}$ & 1.76\%  & 8.47\% & 50\%  \\
    
    {}&{\textbf{48 dB}} & {$3.35\cdot10^{-10}$}&{}  & $\textit{\textbf{d}}$  & $2.01\cdot10^{-8}$ & $1.55\cdot10^{-9}$  & $1.29\cdot10^{-10}$ & {} & $\textit{\textbf{d}}$ &  8.88\%   & 10.34\%  & 50\% \\  
    
    {}& {}&{} &{}  & $\textit{\textbf{v}}$  & $3.11\cdot10^{-9}$ & $1.71\cdot10^{-10}$  & $2.31\cdot10^{-12}$ & {} & $\textit{\textbf{v}}$  & 50\%  & 50\%  & 50\%   \\[0.2cm] 
    \hline
    
    {}& {} &{} &{} & {} &   $\textit{\textbf{s}}$ & $\textit{\textbf{d}}$ & $\textit{\textbf{v}}$ & {} &{}& $\textit{\textbf{s}}$ & $\textit{\textbf{d}}$ & $\textit{\textbf{v}}$ \\
    
    {}& {} &{} &{} & $\textit{\textbf{s}}$  & $8.96\cdot10^{-5}$ & $5.99\cdot10^{-6}$  & $4.45\cdot10^{-7}$ & {} & $\textit{\textbf{s}}$ & 0.727\%  & 4.01\% & 50\%  \\ 
    
   {}&{\textbf{26 dB}} & {$3.84\cdot10^{-8}$} &{}  & $\textit{\textbf{d}}$  & $6.30\cdot10^{-6}$ & $3.84\cdot10^{-7}$  & $6.46\cdot10^{-9}$ & {} & $\textit{\textbf{d}}$ & 3.04\%  & 1.88\% & 50\% \\ 
    
    {}& {}&{} &{}  & $\textit{\textbf{v}}$  & $3.57\cdot10^{-7}$ & $6.19\cdot10^{-9}$  & $1.61\cdot10^{-10}$ & {} & $\textit{\textbf{v}}$  & 50\%   & 50\%  & 50\% \\[0.2cm]  
        
    {}& {}&{} &{}  & {} &  $\textit{\textbf{s}}$ & $\textit{\textbf{d}}$ & $\textit{\textbf{v}}$ & {} &{}& $\textit{\textbf{s}}$ & $\textit{\textbf{d}}$ & $\textit{\textbf{v}}$ \\
    
    {}& {}&{} &{}  & $\textit{\textbf{s}}$  & $1.18\cdot10^{-5}$ & $7.97\cdot10^{-7}$  & $6.12\cdot10^{-8}$ & {} & $\textit{\textbf{s}}$ & 0.841\%  & 4.10\% & 50\%  \\  
    
     {\textbf{Deployed}}&{\textbf{35 dB}} & {$2.07\cdot10^{-9}$}&{}  & $\textit{\textbf{d}}$  & $8.19\cdot10^{-7}$ & $5.03\cdot10^{-8}$  & $1.09\cdot10^{-9}$ & {} & $\textit{\textbf{d}}$ & 3.04\%  & 1.88\% & 50\% \\ 

    {}& {} &{} &{} & $\textit{\textbf{v}}$  & $5.17\cdot10^{-8}$ & $9.71\cdot10^{-10}$  & $8.59\cdot10^{-12}$ & {} & $\textit{\textbf{v}}$  & 50\%   & 50\%  &50\%   \\[0.2cm]
        
    {}& {} &{} &{} & {} &   $\textit{\textbf{s}}$ & $\textit{\textbf{d}}$ & $\textit{\textbf{v}}$ & {} &{}& $\textit{\textbf{s}}$ & $\textit{\textbf{d}}$ & $\textit{\textbf{v}}$ \\
    
    {}& {} &{} &{} & $\textit{\textbf{s}}$  & $1.68\cdot10^{-6}$ & $1.15\cdot10^{-7}$  & $9.79\cdot10^{-9}$ & {} & $\textit{\textbf{s}}$ & 0.837\%  & 4.65\% & 50\%  \\
    
    {}&{\textbf{44 dB}} & {$2.29\cdot10^{-10}$} &{} & $\textit{\textbf{d}}$  & $1.20\cdot10^{-7}$ & $7.34\cdot10^{-9}$  & $3.07\cdot10^{-10}$ & {} & $\textit{\textbf{d}}$ & 4.09\%  & 4.51\% & 50\% \\ 
    
    {}& {} &{} &{} & $\textit{\textbf{v}}$  & $9.04\cdot10^{-9}$ & $3.01\cdot10^{-10}$  & $3.99\cdot10^{-12}$ & {} & $\textit{\textbf{v}}$  & 50\%  & 50\%  & 50\%   \\[0.2cm]
\end{tabular}
\end{table*}
\clearpage

\begin{table}[]
\caption{Measured X-basis Gains and QBERs for the launch powers in the experiment, for both lab and deployed environments.}
\label{tab:gains-qbers-X-launch}
\begin{tabular}{cccccccccccc}
\multicolumn{12}{c}{\textbf{X-basis gains and QBERs}} \\ \hline \hline
                                                     & \textbf{Launch Power} & \textbf{Keyrate}  &  & &  \multicolumn{3}{c}{\textbf{Gain, $Q_{\mu_{i}\mu_{j}}^{X}$}}                                      & \multicolumn{1}{c}{\textbf{}} & \multicolumn{3}{c}{\textbf{QBER, $e_{\mu_{i},\mu_{j}}^{X}$}}                                     \\ \cline{3-12} 
                                                     &                       &                    &  &                                             &               \textbf{$\mu_{b} \rightarrow$}             &                               &                               &                                &                                &                                \\
\multirow{24}{*}{\textbf{\begin{tabular}[c]{@{}c@{}}Lab \\ (19.5 dB)\end{tabular}}}             &                       &  & \multicolumn{1}{r}{\textbf{$\mu_{a} \downarrow$}}  & & \multicolumn{1}{c}{ \textit{\textbf{s}}} & \multicolumn{1}{c}{\textbf{d}} & \multicolumn{1}{c}{ \textit{\textbf{v}}} & \multicolumn{1}{c}{\textbf{}} & \multicolumn{1}{c}{ \textit{\textbf{s}}} & \multicolumn{1}{c}{ \textit{\textbf{d}}} & \multicolumn{1}{c}{ \textit{\textbf{v}}} \\
                                                     &                       &        &         &  \textit{\textbf{s}}                                        & $4.94\cdot10^{-4}$             & $1.12\cdot10^{-4}$             & $9.51\cdot10^{-5}$             &  \textit{\textbf{s}}                    & 29.5\%                         & 43.9\%                         & 50\%                           \\
                                                     & \textbf{4.68 $\mu$W}  & $2.10\cdot10^{-7}$  & &  \textit{\textbf{d}}                                        & $1.75\cdot10^{-4}$             & $2.32\cdot10^{-6}$             & $4.81\cdot10^{-7}$             &  \textit{\textbf{d}}                    & 46.5\%                         & 32.1\%                         & 50\%                           \\ 
                                                     &                       &                &  &  \textit{\textbf{v}}                                        & $1.60\cdot10^{-4}$             & $8.24\cdot10^{-7}$             & $1.08\cdot10^{-9}$             &  \textit{\textbf{v}}                    & 50\%                           & 50\%                           & 50\%                           \\ [ 0.2cm]
                                                     &                       &                 &   & \multicolumn{1}{r}{\textbf{}}                     & \multicolumn{1}{c}{ \textit{\textbf{s}}} & \multicolumn{1}{c}{ \textit{\textbf{d}}} & \multicolumn{1}{c}{ \textit{\textbf{v}}} & \multicolumn{1}{c}{\textbf{}} & \multicolumn{1}{c}{ \textit{\textbf{s}}} & \multicolumn{1}{c}{ \textit{\textbf{d}}} & \multicolumn{1}{c}{ \textit{\textbf{v}}} \\
                                                     &                       &                &  &  \textit{\textbf{s}}                                        & $4.69\cdot10^{-4}$             & $1.08\cdot10^{-4}$             & $9.09\cdot10^{-5}$             &  \textit{\textbf{s}}                    & 29.2\%                         & 43.7\%                         & 50\%                           \\
                                                     & \textbf{11.8 $\mu$W}  & $2.29\cdot10^{-7}$ & &  \textit{\textbf{d}}                                        & $1.67\cdot10^{-4}$             & $2.30\cdot10^{-6}$             & $4.72\cdot10^{-7}$             &  \textit{\textbf{d}}                    & 46.4\%                         & 31.7\%                         & 50\%                           \\ 
                                                     &                       &                 &  &  \textit{\textbf{v}}                                        & $1.52\cdot10^{-4}$             & $8.35\cdot10^{-7}$             & $7.32\cdot10^{-10}$            &  \textit{\textbf{v}}                    & 50\%                           & 50\%                           & 50\%                           \\[ 0.2cm]
                                                     &                       &                  &  \multicolumn{1}{r}{\textbf{}}                     & \multicolumn{1}{c}{ \textit{\textbf{s}}} & \multicolumn{1}{c}{ \textit{\textbf{d}}} & \multicolumn{1}{c}{ \textit{\textbf{v}}} & \multicolumn{1}{c}{\textbf{}} & \multicolumn{1}{c}{ \textit{\textbf{s}}} & \multicolumn{1}{c}{ \textit{\textbf{d}}} & \multicolumn{1}{c}{ \textit{\textbf{v}}} \\
                                                     &                       &          &   &  \textit{\textbf{s}}                                        & $4.81\cdot10^{-4}$             & $1.09\cdot10^{-4}$             & $9.22\cdot10^{-5}$             &  \textit{\textbf{s}}                    & 29.3\%                         & 43.7\%                         & 50\%                           \\
                                                     & \textbf{15.0 $\mu$W}  & $2.09\cdot10^{-7}$  & &  \textit{\textbf{d}}                                        & $1.72\cdot10^{-4}$             & $2.29\cdot10^{-6}$             & $4.67\cdot10^{-7}$             &  \textit{\textbf{d}}                    & 46.6\%                         & 32.2\%                         & 50\%                           \\ 
                                                     &                       &                &  &  \textit{\textbf{v}}                                        & $1.57\cdot10^{-4}$             & $8.37\cdot10^{-7}$             & $1.48\cdot10^{-9}$             &  \textit{\textbf{v}}                    & 50\%                           & 50\%                           & 50\%                           \\[ 0.2cm]
                                                     &                       &               &   & \multicolumn{1}{r}{\textbf{}}                     & \multicolumn{1}{c}{ \textit{\textbf{s}}} & \multicolumn{1}{c}{ \textit{\textbf{d}}} & \multicolumn{1}{c}{ \textit{\textbf{v}}} & \multicolumn{1}{c}{\textbf{}} & \multicolumn{1}{c}{ \textit{\textbf{s}}} & \multicolumn{1}{c}{ \textit{\textbf{d}}} & \multicolumn{1}{c}{ \textit{\textbf{v}}} \\
                                                     &                       &                  & &  \textit{\textbf{s}}                                        & $4.57\cdot10^{-4}$             & $1.08\cdot10^{-4}$             & $9.14\cdot10^{-5}$             &  \textit{\textbf{s}}                    & 29.3\%                      & 43.9\%                     & 50\%                           \\
                                                     & \textbf{61.7 $\mu$W}  & $1.95\cdot10^{-7}$ & &  \textit{\textbf{d}}                                        & $1.58\cdot10^{-4}$             & $2.25\cdot10^{-6}$             & $4.74\cdot10^{-7}$             &  \textit{\textbf{d}}                    & 46.4\%                      & 32.6\%                        & 50\%                           \\
                                                     &                       &            &  &  \textit{\textbf{v}}                                        & $1.44\cdot10^{-4}$             & $8.24\cdot10^{-7}$             & $2.60\cdot10^{-9}$             &  \textit{\textbf{v}}                    & 50\%                           & 50\%                           & 50\%                           \\ [ 0.2cm]
                                                     &                       &           &  & \multicolumn{1}{r}{}                              & \multicolumn{1}{c}{ \textit{\textbf{s}}} & \multicolumn{1}{c}{ \textit{\textbf{d}}} & \multicolumn{1}{c}{ \textit{\textbf{v}}} & \multicolumn{1}{c}{\textbf{}} & \multicolumn{1}{c}{ \textit{\textbf{s}}} & \multicolumn{1}{c}{ \textit{\textbf{d}}} & \multicolumn{1}{c}{ \textit{\textbf{v}}} \\
                                                     &                       &        &    &  \textit{\textbf{s}}                                        & $4.56\cdot10^{-4}$             & $1.08\cdot10^{-4}$             & $9.11\cdot10^{-5}$             &  \textit{\textbf{s}}                    & 29.3\%                         & 43.9\%                         & 50\%                           \\
                                                     & \textbf{155 $\mu$W}   & $9.70\cdot10^{-8}$ & &  \textit{\textbf{d}}                                        & $1.60\cdot10^{-4}$             & $2.36\cdot10^{-6}$             & $5.10\cdot10^{-7}$             &  \textit{\textbf{d}}                    & 46.4\%                         & 32.6\%                         & 50\%                           \\ 
                                                     &                       &           &     &     \textit{\textbf{v}}                                        & $1.46\cdot10^{-4}$             & $9.01\cdot10^{-7}$             & $8.16\cdot10^{-9}$             &  \textit{\textbf{v}}                    & 50\%                           & 50\%                           & 50\%                           \\[ 0.2cm]
                                                     &                       &          &      &    \multicolumn{1}{r}{}                              & \multicolumn{1}{c}{ \textit{\textbf{s}}} & \multicolumn{1}{c}{ \textit{\textbf{d}}} & \multicolumn{1}{c}{ \textit{\textbf{v}}} & \multicolumn{1}{c}{\textbf{}} & \multicolumn{1}{c}{ \textit{\textbf{s}}} & \multicolumn{1}{c}{ \textit{\textbf{d}}} & \multicolumn{1}{c}{ \textit{\textbf{v}}} \\
                                                     &                       &          &        &    \textit{\textbf{s}}                                        & $4.67\cdot10^{-4}$             & $1.07\cdot10^{-4}$             & $8.94\cdot10^{-5}$             &  \textit{\textbf{s}}                    & 29.4\%                         & 43.9\%                         & 50\%                           \\
                                                     & \textbf{392 $\mu$W}   & $1.02\cdot10^{-8}$ & &  \textit{\textbf{d}}                                        & $1.70\cdot10^{-4}$             & $2.72\cdot10^{-6}$             & $6.35\cdot10^{-7}$             &  \textit{\textbf{d}}                    & 46.5\%                         & 33.4\%                         & 50\%                           \\ 
                                                     &                       &          &       &     \textit{\textbf{v}}                                        & $1.55\cdot10^{-4}$             & $1.05\cdot10^{-6}$             & $1.86\cdot10^{-8}$             &  \textit{\textbf{v}}                    & 50\%                           & 50\%                           & 50\%                           \\ \hline
                                                     &                       &         &         &   \multicolumn{1}{r}{}                              & \multicolumn{1}{c}{ \textit{\textbf{s}}} & \multicolumn{1}{c}{ \textit{\textbf{d}}} & \multicolumn{1}{c}{ \textit{\textbf{v}}} & \multicolumn{1}{c}{\textbf{}} & \multicolumn{1}{c}{ \textit{\textbf{s}}} & \multicolumn{1}{c}{ \textit{\textbf{d}}} & \multicolumn{1}{c}{ \textit{\textbf{v}}} \\
\multirow{4}{*}{\textbf{\begin{tabular}[c]{@{}c@{}}Deployed\\  (26 dB)\end{tabular}}}                                           &                       &        &         &     \textit{\textbf{s}}                                        & $1.72\cdot10^{-4}$             & $5.53\cdot10^{-5}$             & $4.99\cdot10^{-5}$             &  \textit{\textbf{s}}                    & 29.9\%                         & 46.22\%                        & 50\%                           \\
                           & \textbf{10.8 $\mu$W}  & $3.84\cdot10^{-8}$ & &  \textit{\textbf{d}}                                        & $4.40\cdot10^{-5}$             & $8.01\cdot10^{-7}$             & $2.53\cdot10^{-7}$             &  \textit{\textbf{d}}                    & 44.6\%                         & 32.0\%                         & 50\%                           \\
\textbf{}                                            &                       &          &     &       \textit{\textbf{v}}                                        & $3.78\cdot10^{-5}$             & $1.76\cdot10^{-7}$             & $1.28\cdot10^{-10}$            &  \textit{\textbf{v}}                    & 50\%                           & 50\%                           & 50\%                          
\end{tabular}
\end{table}

\clearpage

\begin{table}[]
\caption{Measured Z-basis Gains and QBERs for the launch powers in the experiment, for both lab and deployed environments.}
\label{tab:gains-qbers-Z-launch}
\begin{tabular}{ccclcccccccc}
\multicolumn{12}{c}{\textbf{Z-basis gains and QBERs}}                                                                                                                                                                                                                                                                                                         \\ \hline \hline
                                                                                      & \textbf{Launch Power} & \textbf{Keyrate}   & \multicolumn{1}{c}{}          &                     & \multicolumn{3}{c}{\textbf{Gain, $Q_{\mu_{i}\mu_{j}}^{Z}$}}                & \textbf{}           & \multicolumn{3}{c}{\textbf{QBER, $e_{\mu_{i},\mu_{j}}^{Z}$}}    \\ \cline{3-12} 
                                                                                      &                       &                    &                               &                     & \textbf{$\mu_{b} \rightarrow$} &                     &                     &                     &                     &                     &                     \\
\multirow{24}{*}{\textbf{\begin{tabular}[c]{@{}c@{}}Lab \\ (19.5 dB)\end{tabular}}}   &                       &                    & \textbf{$\mu_{a} \downarrow$} & \textbf{}           & \textit{\textbf{s}}            & \textit{\textbf{d}} & \textit{\textbf{v}} & \textit{\textbf{}}  & \textit{\textbf{s}} & \textit{\textbf{d}} & \textit{\textbf{v}} \\
                                                                                      &                       &                    &                               & \textit{\textbf{s}} & $2.33\cdot10^{-4}$             & $1.68\cdot10^{-5}$  & $1.40\cdot10^{-6}$  & \textit{\textbf{s}} & 0.927\%             & 4.54\%              & 50\%                \\
                                                                                      & \textbf{4.68 $\mu$W}  & $2.10\cdot10^{-7}$ &                               & \textit{\textbf{d}} & $1.60\cdot10^{-5}$             & $1.07\cdot10^{-6}$  & $4.65\cdot10^{-8}$  & \textit{\textbf{d}} & 5.37\%              & 5.20\%              & 50\%                \\
                                                                                      &                       &                    &                               & \textit{\textbf{v}} & $1.60\cdot10^{-6}$             & $6.18\cdot10^{-8}$  & $1.04\cdot10^{-9}$  & \textit{\textbf{v}} & 50\%                & 50\%                & 50\%                \\[0.2cm]
                                                                                      &                       &                    &                               & \textit{\textbf{}}  & \textit{\textbf{s}}            & \textit{\textbf{d}} & \textit{\textbf{v}} & \textbf{}           & \textit{\textbf{s}} & \textit{\textbf{d}} & \textit{\textbf{v}} \\
                                                                                      &                       &                    &                               & \textit{\textbf{s}} & $2.34\cdot10^{-4}$             & $1.71\cdot10^{-5}$  & $1.46\cdot10^{-6}$  & \textit{\textbf{s}} & 1.14\%              & 4.74\%              & 50\%                \\
                                                                                      & \textbf{11.8 $\mu$W}  & $2.29\cdot10^{-7}$ &                               & \textit{\textbf{d}} & $1.65\cdot10^{-5}$             & $1.10\cdot10^{-6}$  & $4.80\cdot10^{-8}$  & \textit{\textbf{d}} & 6.31\%              & 5.50\%              & 50\%                \\
                                                                                      &                       &                    &                               & \textit{\textbf{v}} & $1.92\cdot10^{-6}$             & $6.66\cdot10^{-8}$  & $1.06\cdot10^{-9}$  & \textit{\textbf{v}} & 50\%                & 50\%                & 50\%                \\[0.2cm]
                                                                                      &                       &                    &                               & \textit{\textbf{}}  & \textbf{s}                     & \textbf{d}          & \textbf{v}          & \textit{\textbf{}}  & \textbf{s}          & \textbf{d}          & \textbf{v}          \\
                                                                                      &                       &                    &                               & \textit{\textbf{s}} & $2.38\cdot10^{-4}$             & $1.73\cdot10^{-5}$  & $1.41\cdot10^{-6}$  & \textit{\textbf{s}} & 1.00\%              & 4.50\%              & 50\%                \\
                                                                                      & \textbf{15.0 $\mu$W}  & $2.09\cdot10^{-7}$ &                               & \textit{\textbf{d}} & $1.64\cdot10^{-5}$             & $1.11\cdot10^{-6}$  & $5.29\cdot10^{-8}$  & \textit{\textbf{d}} & 6.28\%              & 5.86\%              & 50\%                \\
                                                                                      &                       &                    &                               & \textit{\textbf{v}} & $1.95\cdot10^{-6}$             & $7.72\cdot10^{-8}$  & $1.61\cdot10^{-9}$  & \textit{\textbf{v}} & 50\%                & 50\%                & 50\%                \\[0.2cm]
                                                                                      &                       &                    &                               & \textit{\textbf{}}  & \textbf{s}                     & \textbf{d}          & \textbf{v}          & \textit{\textbf{}}  & \textbf{s}          & \textbf{d}          & \textbf{v}          \\
                                                                                      &                       &                    &                               & \textit{\textbf{s}} & $2.32\cdot10^{-4}$             & $1.73\cdot10^{-5}$  & $1.57\cdot10^{-6}$  & \textit{\textbf{s}} & 1.03\%              & 4.95\%              & 50\%                \\
                                                                                      & \textbf{61.7 $\mu$W}  & $1.95\cdot10^{-7}$ &                               & \textit{\textbf{d}} & $1.63\cdot10^{-5}$             & $1.12\cdot10^{-6}$  & $6.77\cdot10^{-8}$  & \textit{\textbf{d}} & 6.92\%              & 7.24\%              & 50\%                \\
                                                                                      &                       &                    &                               & \textit{\textbf{v}} & $2.13\cdot10^{-6}$             & $9.40\cdot10^{-8}$  & $2.70\cdot10^{-9}$  & \textit{\textbf{v}} & 50\%                & 50\%                & 50\%                \\[0.2cm]
                                                                                      &                       &                    &                               & \textit{}           & \textbf{s}                     & \textbf{d}          & \textbf{v}          & \textit{\textbf{}}  & \textbf{s}          & \textbf{d}          & \textbf{v}          \\
                                                                                      &                       &                    &                               & \textit{\textbf{s}} & $2.40\cdot10^{-4}$             & $1.83\cdot10^{-5}$  & $2.23\cdot10^{-6}$  & \textit{\textbf{s}} & 1.35\%              & 6.64\%              & 50\%                \\
                                                                                      & \textbf{155 $\mu$W}   & $9.70\cdot10^{-8}$ &                               & \textit{\textbf{d}} & $1.77\cdot10^{-5}$             & $1.25\cdot10^{-6}$  & $1.16\cdot10^{-8}$  & \textit{\textbf{d}} & 9.51\%              & 11.0\%              & 50\%                \\
                                                                                      &                       &                    &                               & \textit{\textbf{v}} & $3.20\cdot10^{-6}$             & $1.66\cdot10^{-7}$  & $8.27\cdot10^{-9}$  & \textit{\textbf{v}} & 50\%                & 50\%                & 50\%                \\[0.2cm]
                                                                                      &                       &                    &                               & \textit{}           & \textbf{s}                     & \textbf{d}          & \textbf{v}          & \textit{\textbf{}}  & \textbf{s}          & \textbf{d}          & \textbf{v}          \\
                                                                                      &                       &                    &                               & \textit{\textbf{s}} & $2.35\cdot10^{-4}$             & $1.95\cdot10^{-5}$  & $3.96\cdot10^{-6}$  & \textit{\textbf{s}} & 2.38\%              & 11.1\%              & 50\%                \\
                                                                                      & \textbf{392 $\mu$W}   & $1.02\cdot10^{-8}$ &                               & \textit{\textbf{d}} & $1.65\cdot10^{-5}$             & $1.43\cdot10^{-6}$  & $1.94\cdot10^{-7}$  & \textit{\textbf{d}} & 12.5\%              & 15.2\%              & 50\%                \\
                                                                                      &                       &                    &                               & \textit{\textbf{v}} & $4.52\cdot10^{-6}$             & $2.50\cdot10^{-8}$  & $1.87\cdot10^{-8}$  & \textit{\textbf{v}} & 50\%                & 50\%                & 50\%                \\[0.2cm]\hline
\multirow{4}{*}{\textbf{\begin{tabular}[c]{@{}c@{}}Deployed\\  (26 dB)\end{tabular}}} &                       &                    &                               &                     & \textit{\textbf{s}}            & \textit{\textbf{d}} & \textit{\textbf{v}} & \textbf{}           & \textit{\textbf{s}} & \textit{\textbf{d}} & \textit{\textbf{v}} \\
                                                                                      &                       &                    &                               & \textit{\textbf{s}} & $8.96\cdot10^{-5}$             & $5.99\cdot10^{-6}$  & $4.45\cdot10^{-7}$  & \textit{\textbf{s}} & 0.728\%             & 4.01\%              & 50\%                \\
                                                                                      & \textbf{10.8 $\mu$W}  & $3.84\cdot10^{-8}$ &                               & \textit{\textbf{d}} & $6.30\cdot10^{-6}$             & $3.84\cdot10^{-7}$  & $6.46\cdot10^{-9}$  & \textit{\textbf{d}} & 3.04\%              & 1.88\%              & 50\%                \\
                                                                                      &                       &                    &                               & \textit{\textbf{v}} & $3.57\cdot10^{-7}$             & $6.19\cdot10^{-9}$  & $1.61\cdot10^{-10}$ & \textit{\textbf{v}} & 50\%                & 50\%                & 50\%               
\end{tabular}
\end{table}

\section{Supplementary B: Qubit parameters} \label{sec:AppendixB}

We acquired the numerical simulation results using parameters that describe the quality of the qubits that our system generates. Each qubit state is generally described by Eq.~\ref{eq:appqubitmodel} where the $m$ and $\phi$ parameters of the various qubit states can be found in table~\ref{tab:qubit-params} for both Alice and Bob, for both both lab and deployed settings.

\begin{table*}[h]
\centering
\caption{Qubit parameters according to the model in Eq.~\ref{eq:appqubitmodel}, which were used to create the simulation graphs in figures \ref{fig:dataset2} and \ref{fig:dataset1}. The visibility in the lab setting was set to 0.847, and in the deployed setting to 0.797.}
\label{tab:qubit-params}
\begin{tabular}{llllllllll}
\multicolumn{10}{c}{\textbf{Qubit parameters}}                                                                                                                                                                                                                                                                                                                                                    \\ \hline \hline
                                   &                      & \multicolumn{4}{c}{\textbf{Alice}}                                                                                                                                & \multicolumn{4}{c}{\textbf{Bob}}                                                                                                                                  \\ \cline{3-10} 
                                   & \multicolumn{1}{r}{} & \multicolumn{1}{c}{\textbf{$\ket{e}$}} & \multicolumn{1}{c}{\textbf{$\ket{l}$}} & \multicolumn{1}{c}{\textbf{$\ket{+}$}} & \multicolumn{1}{c}{\textbf{$\ket{-}$}} & \multicolumn{1}{c}{\textbf{$\ket{e}$}} & \multicolumn{1}{c}{\textbf{$\ket{l}$}} & \multicolumn{1}{c}{\textbf{$\ket{+}$}} & \multicolumn{1}{c}{\textbf{$\ket{-}$}} \\
\multirow{6}{*}{\textbf{Lab}}      & \textbf{$m_s$}       & 0.9950                                  & 0.0045                                 & 0.5156                                 & 0.5260                                 & 0.9989                                 & 0.0005                                 & 0.5255                                 & 0.5243                                 \\
                                   & \textbf{$\phi_s$}    & 0                                      & 0                                      & 0                                      & 3.14                                  & 0                                      & 0                                      & 0                                      & 3.2                                  \\
                                   & \textbf{$m_d$}       & 0.9524                                 & 0.0403                                      & 0.4937                                 & 0.5040                                 & 0.9577                                 & 0.0392                                      & 0.5077                                 & 0.5060                                 \\
                                   & \textbf{$\phi_d$}    & 0                                      & 0                                      & 0                                      & 3.54                               & 0                                      & 0                                      & 0                                      & 2.78                               \\ 
                                   \hline
\multirow{6}{*}{\textbf{Deployed}} & \textbf{$m_s$}       & 0.9969                                 & 0.002                                  & 0.5212                                 & 0.5203                                 & 0.9968                                 & 0.0015                                 & 0.4984                                 & 0.4963                                 \\
                                   & \textbf{$\phi_s$}    & 0                                      & 0                                      & 0                                      & 3.14                                  & 0                                      & 0                                      & 0                                      & 3.2                                  \\
                                   & \textbf{$m_d$}       & 0.9612                                 & 0.0055                                 & 0.4904                                 & 0.4911                                 & 0.964                                  & 0.0006                                 & 0.4888                                 & 0.4857                                 \\
                                   & \textbf{$\phi_d$}    & 0                                      & 0                                      & 0                                      & 3.48                               & 0                                      & 0                                      & 0                                      & 2.80                              
\end{tabular}
\end{table*}

\end{document}